\documentclass[conference]{IEEEtran}

\IEEEoverridecommandlockouts

\newif\ifdoc

\usepackage{empheq}
\usepackage{amsmath,amssymb}
\usepackage{relsize}

\usepackage[mathscr]{euscript}
\usepackage{color,hyperref}
\usepackage[normalem]{ulem}
\usepackage{indentfirst}
\usepackage{algorithm}
\usepackage{algorithmicx}
\usepackage{algpseudocode}

\ifdoc
\renewcommand{\mathbb}{{}}
\renewcommand{\boldsymbol}[1]{{#1}}
\renewcommand{\mathcal}[1]{{#1}}
\renewcommand{\binom}[2]{{#1\choose#2}}
\renewcommand{\operatorname}[1]{{#1}}
\fi

\newcommand{\F}{\mathbb F}
\newcommand{\N}{\mathbb N}
\newcommand{\ev}{\operatorname{ev}}
\newcommand{\set}[1]{\left\{#1\right\}}
\newcommand{\size}[1]{\left|#1\right|}

\renewcommand{\refeq}[1]{(\ref{#1})}

\newcommand{\MRM}{\text{Mult}}
\newcommand{\RM}{\text{RM}}

\newcommand{\Enc}{\text{Enc}}

\newcommand{\Hasse}[2]{H(#1,#2)}

\newcommand{\vX}{{\boldsymbol X}}

\newcommand{\vZ}{{\boldsymbol Z}}

\newcommand{\vi}{{\boldsymbol i}}
\newcommand{\vj}{{\boldsymbol j}}
\newcommand{\vk}{{\boldsymbol k}}

\newcommand{\vu}{{\boldsymbol u}}
\newcommand{\vv}{{\boldsymbol v}}

\newcommand{\vP}{{\boldsymbol P}}

\newcommand{\Infoset}{\mathcal I}

\newtheorem{proposition}{Proposition}
\newtheorem{definition}{Definition}
\newtheorem{theorem}{Theorem}
\newtheorem{lemma}{Lemma}

\begin{document}

\title{Information Sets of Multiplicity Codes}
\author{\IEEEauthorblockN{Daniel Augot}
\IEEEauthorblockA{INRIA Saclay and LIX\\
B\^atiment Alan Turing \\
1 rue Honor\'e d'Estienne d'Orves\\
91120 Palaiseau\\
\href{mailto:daniel.augot@inria}{daniel.augot@inria.fr}}
\and
\IEEEauthorblockN{Fran\c{c}oise Levy-dit-Vehel}
\IEEEauthorblockA{ENSTA ParisTech\\
828 boulevard des Mar\'echaux\\
91762 Palaiseau\\
INRIA Saclay and LIX\\
\href{mailto:levy@ensta.fr}{levy@ensta.fr}}
\and
\IEEEauthorblockN{Cuong M. Ng\^o\thanks{Ng\^o is supported by INRIA and Alcatel-Lucent \emph{Privacy} ADR}}
\IEEEauthorblockA{INRIA Saclay and LIX\\
B\^atiment Alan Turing\\
1 rue Honor\'e d'Estienne d'Orves\\
91120 Palaiseau\\
\href{mailto:manh-cuong.ngo@inria.fr}{manh-cuong.ngo@inria.fr}}}

\maketitle
\begin{abstract}
  We here provide a method for systematic encoding of the Multiplicity codes 
introduced by Kopparty, Saraf and Yekhanin in 2011. The construction is built 
on an idea of Kopparty. We properly define information sets for these codes and give detailed proofs of the validity of Kopparty's construction, that use generating functions. We also give a complexity estimate of the associated
  encoding algorithm.
\end{abstract}
\section{Introduction}
Locally decodable codes (LDC) allow one to probabilistically retrieve one
symbol of a \emph{message} by looking at only a small fraction of its
encoding. They were formally introduced by Katz and Trevisan in
2000~\cite{KT00}. When the local decoding algorithm retrieves a symbol
of the \emph{codeword} instead of a message symbol, one speaks of locally
correctable codes (LCC). For an extensive treatment of locally
decodable and correctable codes, we refer the reader to~\cite{YEK10}.




For $C$ to be an LCC code, it is only required to have $C$ defined as
$C\subset\F_q^n$, while the notion of an LDC code requires that $C$ is
provided with an encoding $\Enc:\F_q^k\rightarrow\F_q^n$.  Considering
codes which are $\F_q$-linear subspaces of $\F_q^n$, there is a
reduction making an LDC code from an LCC code~\cite[Lemma
2.3]{YEK10}. This reduction heavily relies on the notion of
\emph{Information Set}.

A breakthrough of Kopparty, Saraf and Yekhanin~\cite{KSY11} is a
construction of high-rate LCCs with sublinear locality. These codes
were termed {\em Multiplicity Codes}, and generalize the Reed-Muller
codes, using derivatives.

A technical and practical issue remains, which is to make these codes
LDCs. For these codes, the message space and the codeword space do not
share the same alphabet, so the standard reduction from~\cite[Lemma
2.3]{YEK10} can not be applied. The problem was circumvented
in~\cite{KSY11} by using concatenation.

It is well known that LDCs can be used to build Private Information
Retrieval (PIR) schemes, using a standard equivalence between LDCs and
PIRs~\cite{KT00}.  In~\cite{AFS2014}, for the very particular case of
Reed-Muller codes and Multiplicity codes, a better usage of these
locally decodable codes in PIR schemes was introduced, using a
partitioning of the $m$-dimensional affine space into few affine
hyperplanes. The concatenation solution provided by~\cite{KSY11}
appears not helpful in this context, since it more or less breaks the
underlying affine geometry.

In the Appendix of~\cite{KOP12}, Kopparty described an idea to make a
systematic encoding for Multiplicity codes. We clarify the idea
of~\cite{KOP12}, providing notation and proofs, and solve a unicity
problem, necessary to have a valid systematic encoding.

\section{Problem statement}
Let $q = p^t$ for some $t \in \mathbb{N}^{*}$ and $p$ prime. We
enumerate the field with $q$ elements as $\mathbb{F}_q = \lbrace
\alpha_0,\alpha_1,\dots,\alpha_{q-1} \rbrace$. Considering, for $m \in
\N^*$, $m$ indeterminates $X_1,\dots,X_m$ and $m$ positive integers
$i_1,\dots,i_m$, we use the short-hand notation
\begin{align*}
\vX&=(X_1,\dots,X_m) \quad \quad \vX^\vi= X_1^{i_1}\cdots X_m^{i_m},\\
\F_q[\vX]&=\F_q[X_1,\dots,X_m] \quad \quad \vi= (i_1,\dots,i_m)\in\N^m,\\
|\vi|&=i_1+\dots+i_m \quad \quad \vP=(p_1,\dots,p_m)\in\F_q^m,
\end{align*}
i.e. we use bold symbols for vectors, points, etc, and standard
symbols for uni-dimensionnal scalars, variables, etc. We denote by
\[
\F_q[\vX]_d=\set{F\in\F_q[\vX];\ \deg F\leq d}.
\]
We also let $V = \mathbb{F}_q^m = \lbrace \vP_1,\dots,\vP_n \rbrace$,
where $n=q^m$.
\subsection{Reed-Muller codes over  $\mathbb{F}_q$ and information sets}
We define the following evaluation map
\[
\ev:\begin{array}[t]{rcl}
\F_q[\vX] &\rightarrow&\F_q^n\\
F & \mapsto & \left(F(\vP_1),\dots,F(\vP_n)\right).
\end{array}
\]
For an integer $d>0$, we denote by $\F_q[\vX]_d$ the set of
polynomials of degree less than or equal to $d$, which has dimension
$\binom{m+d}{m}$ over $\F_q$, see~\cite{KOP12}.  We can now recall the
definition of Reed-Muller codes over $\mathbb{F}_q$, also called
Generalized Reed-Muller codes~\cite{KLP68}:
\begin{definition}[Reed-Muller codes over $\mathbb{F}_q$] 
For $d \leq m(q-1)$, the $d^{th}-$ order \emph{Reed-Muller code over} $\mathbb{F}_q$, $\RM_d$ is
$$\RM_d=\left\{\ev(F)\;\mid  F\in \F_q[\vX]_d\right\}.$$
\end{definition}
From now on, we omit ``over $\mathbb{F}_q$'' and simply say Reed-Muller codes.
The evaluation map $\ev$ maps $\binom{m+d}{m}$ symbols into $n$
symbols. However, when $d\geq q$, the map $\ev$ is not injective, and
the dimension $k_d$ of $\RM_d$ is less than or equal to
$\binom{m+d}d$.

A codeword $c\in \RM_d$ can be indexed by integers as
$c=(c_1,\dots,c_n)$ or by points as $c=(c_{\vP_1},\dots,c_{\vP_n})$,
where $c_i=c_{\vP_i}=F(\vP_i)$.

\begin{definition}[Information set]
  Let $\mathcal{C}$ be an $[n,k]$ linear code over $\F_q$. An
  \emph{information set} of $\mathcal{C}$ is a subset $\Infoset \subset
  \lbrace 1,\dots,n \rbrace$ such that the map:
\begin{equation}\label{eq:mapinfoset}
\varphi:\begin{array}[t]{rcl}
\mathcal{C} & \rightarrow &\F_q^k\\
c & \mapsto & (c_i)_{i \in \Infoset}
\end{array}
\end{equation}
is a bijection.
\end{definition}
J.D. Key~\textit{et al.}~\cite{KMM06} gave  information sets for Reed-Muller codes, that we recall in the following Theorem.
\begin{theorem}[\cite{KMM06}]\label{theo:infosetRM} An information set of $\RM_d$ is
\begin{equation*}
\begin{split}
&\left\lbrace (\alpha_{i_1},\dots,\alpha_{i_m}) |\ 0 \leq i_l \leq q-1;\ 1 \leq l \leq m;\text{ }\sum_{l=1}^m i_l \leq d \right\rbrace.
\end{split}
\end{equation*}
We denote this particular information set by $\Infoset_{d} $, with
$\Infoset_d\subset V$.
\end{theorem}
Denote by
\begin{equation*}
\begin{split}
\mathcal{K}_{d} =\lbrace ({i_1},\dots,{i_m})\text{ } |\text{ } 0 \leq i_l \leq q-1,\; 1 \leq l \leq m,\:\sum_{l=1}^m i_l = d \rbrace,
\end{split}
\end{equation*}
\begin{equation*}
\begin{split}
\mathcal{L}_{d} =\lbrace ({i_1},\dots,{i_m})\text{ } |\text{ } 0 \leq i_l \leq q-1,\; 1 \leq l \leq m,\:\sum_{l=1}^m i_l \leq d \rbrace,
\end{split}
\end{equation*}
then we have  $k_d =
\dim(\RM_d) = |\mathcal{L}_d| = \sum_{l=0}^d
|\mathcal{K}_{l}|$ (see~\cite{KLP68}).
\subsection{Multiplicity codes}
First we recall the notion of Hasse derivative for multivariate
polynomials. We write polynomials
$F\in\F_q[\vX]=\F_q[X_1,\dots,X_m]$ without parentheses and without
variables, and $F(\vX)$ (resp. $F(\vP)$) when the evaluation on
indeterminates (resp. points) has to be specified. Given $\vi, \vj \in
\mathbb{N}^m$, we denote:
\begin{itemize}
	\item $\vi \leq \vj$ if $i_l \leq j_l$ for all $l = 1,\dots,m$,
	\item $\vi < \vj$ if $\vi \leq \vj$ and $i_l<j_l$ for some $ 1\leq
          l \leq m$.
\end{itemize}
Given  $\vi\in\N^m$, and $F\in\mathbb F_q[\vX]$, the $\vi$-th
\emph{Hasse derivative} of $F$, denoted by $\Hasse F {\vi}$, is the
coefficient of $\vZ^{\vi}$ in the polynomial $F(\vX+\vZ)
\in\F_q[\vX,\vZ]$, where $\vZ=(Z_1,\ldots,Z_m)$. More specifically,
let $F(\vX)=\sum_{\vj \geq \textbf{0}}f_{\vj}\vX^{\vj}$, then
\begin{gather*}
\begin{aligned}
  F(\vX+\vZ)&=\sum_{\vj}f_{\vj}(\vX+\vZ)^{\vj}=\sum_{\vi}\Hasse{F}{\vi}(\vX)\vZ^{\vi},\\
\end{aligned}
\shortintertext{where}
\begin{aligned}
  \Hasse
  F{\vi}(\vX)&=\sum_{\vj\geq\vi}f_{\vj}\binom{\vj}{\vi}\vX^{\vj-\vi},
\end{aligned}
\shortintertext{with}
\begin{aligned} \binom \vj\vi&=\binom {j_1}{i_1}\cdots\binom{j_m}{i_m}.
\end{aligned}
\end{gather*}
Given $F, G \in \F_q[\vX]$ and $\vi\in\N^m$, we have (Leibniz rule~\cite{YEK10}):
\begin{equation}\label{Leibniz}
H(F\cdot G,\vi)= \sum_{\textbf{0} \leq \vk \leq \vi} H(F,\vk)\cdot H(G,\vi - \vk)
\end{equation}
Now, given a derivation order $s>0$, we introduce an extended notion of
evaluation. For a given $s > 0$, there are $\sigma=\binom{m+s-1}{m}$
Hasse derivatives of a polynomial $F$ : $\Hasse F\vi$, $\vi\in
\N^m$, $\size\vi<s$.  Denote by $S = \set{\vj\in \N^m;\
  \size{\vj}< s}$, and let $\Sigma=\F_q^S$. An element
$x\in\Sigma$ is written as
\[
x=(x_\vj)_{\vj\in S},\quad x_\vj\in\mathbb{F}_q.
\]
We generalize the evaluation map at a point $\vP$:
\[
\ev^s_\vP:\begin{array}[t]{rcl}
  \F_q[\vX]&\rightarrow& \Sigma\\
   F&\mapsto &\left(\Hasse F \vv(\vP)\right)_{\vv \in S}
\end{array}
\]
and the total evaluation rule is
\begin{equation*}
\ev^s:\begin{array}[t]{rcl}
  \F_q[\vX]&\rightarrow& \Sigma^n\\
F&\mapsto
  & \left(\ev^s_{\vP_1}(F),\dots,\ev^s_{\vP_n}(F)\right).
\end{array}
\end{equation*}
\begin{definition}[Multiplicity Codes~\cite{KSY11}]\label{def:mult} Given the above
  evaluation map and a degree $d<sq$, the corresponding
  \emph{Multiplicity code} is
\[
\MRM^s_{d}=\left\{\ev^s(F)\;\mid F\in \F_q[\vX]_d\right\}.
\]
\end{definition}
In the context of~\cite{KSY11} the constraint $d<sq$ is
required to ensure that
$\ev^s$ restricted to $\F_q[\vX]_d$ is injective.

\subsection{Information sets of Multiplicity codes}
The difficulty in defining an information set
for a multiplicity code properly is that the $\F_q$-symbols of the message space are
not the same as the $\F_q^S$-symbols of the codeword space.  Recall
that a codeword $c\in \Sigma^{n}$ can be indexed by points $\vP\in V$:
\[
c=\left(c_{\vP}\right)_{\vP\in V}, \quad c_{\vP}\in \Sigma.
\]
Each $c_\vP$ can be written $ c_\vP = ((c_{\vj})_\vP)_{\vj \in S}$,
hence we can write
\[
c = (c_{\vj,\vP})_{\vj \in S,\vP \in V}.
\]
We can now define information sets of Multiplicity Codes:
\begin{definition}[Information set of a Multiplicity Code] An
  \emph{information set} of $\MRM^s_{d}$ is a set $\Infoset \subset
  S \times \mathbb{F}_q^m$ such that the mapping
$$
\phi:\begin{array}[t]{rcl}
\MRM^s_{d}&\rightarrow&\mathbb{F}_q^{\Infoset}\\
c & \mapsto & (c_{\vj,\vP})_{(\vj,\vP) \in \Infoset}
\end{array}
$$
is bijective.
\end{definition}
In~\cite{KOP12}, an information set $\Infoset$ of $\MRM^s_{d}$ based
on information sets of Reed-Muller codes was suggested, namely, $
\Infoset = (\vj,\Infoset_{d_\vj})_{\vj \in S} $ where
$\Infoset_{d_\vj}$ is the information set of the  $d_\vj$-th order Reed-Muller code as in~\refeq{eq:poldecomp}, where the degree $d_\vj$ is
\[
d_\vj=\min(m(q-1),d-\vj q), \ \vj\in S.
\] 
We  prove that $\Infoset$
is an information set in the next two Sections.

\section{Systematic encoding algorithm}
\subsection{A polynomial decomposition}

Given a multi-index $\vj = (j_1,\dots,j_m)$, let $V_\vj$ be the
polynomial
\[
V_\vj=\prod_{i=1}^m(X_i^q-X_i)^{j_i}.
\]
The following decomposition 
is given
in~\cite{KOP12} without proof. 
\begin{lemma}\label{lemma:existence}
Let $F\in\F_q[\vX]$ have total degree less than or equal to $d$, then $F$ can be written as
\begin{equation}\label{eq:poldecomp}
  F=\sum_{|\vj| \leq d/q}F_\vj\cdot V_\vj,
\end{equation}
for some polynomials $F_\vj\in\F_q[\vX]_{d_{\vj}}$. There also exists
a polynomial $F_{\vj_0}$ where $|{\vj_0}| = \lfloor d/q\rfloor$ and
$\deg(F_{\vj_0}) = d-\lfloor d/q\rfloor q$.
\end{lemma}
\begin{IEEEproof}
We consider a multivariate monomial $X_1^{u_1}\dots X_m^{u_m}$ and write $u_i = t_iq + r_i$ for all $i = 1,\dots,m$. First, we consider just $X_1^{u_1}$:
\begin{itemize}
	\item if $t_1 = 0$, since $r_1 <q$, we do not need to prove anything;
	\item if $t_1 > 0$, we have:
	\begin{equation*}
	\begin{split} 
	X_1^{u_1} &= X_1^{r_1}\cdot ((X_1^q-X_1)+X_1)^{t_1} \\
	&= \mathlarger{\mathlarger{\sum}}_{i=0}^{t_1} \binom{t_1}{i}X_1^{r_1+(t_1-i)}\cdot (X_1^q-X_1)^i
	\end{split}
	\end{equation*}
\end{itemize}
Similarly, we recursively apply the above reduction with
$X_1^{r_1+(t_1-i)}$ where $i = 0,\dots,t_1$, so we finally obtain:
\begin{align*}
X_1^{u_1}&= \mathlarger{\sum}_{i=0}^{t_1^\prime} P_{1,i}(X_1)\cdot (X_1^q-X_1)^i\\
&= P_{1,t_1^\prime}(X_1)\cdot (X_1^q-X_1)^{t_1^\prime} + \sum_{i=0}^{t_1^\prime-1} P_{1,i}(X_1)\cdot (X_1^q-X_1)^i,
\end{align*}
where $\deg(P_{1,i}) \leq q-1$ for $i = 0,\dots,t_1^\prime$, for some $t_1^\prime$. We see that
\[
\deg(P_{1,i}(X_1)\cdot (X_1^q-X_1)^i) < q(i+1) \leq qt_1^\prime < u_1,
\]
for all $i = 0,\dots,t_1^\prime -1$, so the term of degree $u_1 =
\deg(X_1^{u_1})$ belongs to $P_{1,t_1^\prime}(X_1)\cdot
(X_1^q-X_1)^{t_1^\prime}$, hence $\deg(P_{1,t_1^\prime}) = u_1 -
qt_1^\prime = r_1 + q(t_1 - t_1^\prime)$.

Since $0 \leq r_1,$ as $\deg(P_{1,t_1^\prime})\leq q-1$, it follows that $t_1 -
t_1^\prime = 0$, so we have
$\deg(P_{1,t_1})\leq\min(q-1,u_1-qt_1)$. Doing the same thing with the
other variables $X_2,\dots,X_m$, we obtain:
\begin{equation*}
\begin{split}
&X_1^{u_1}\cdots X_m^{u_m}
=\left( \mathlarger{\sum}_{i_1=0}^{t_1} P_{1,i_1}(X_1)\cdot (X_1^q-X_1)^{i_1} \right) \cdots \\
&\quad \cdots \left( \mathlarger{\sum}_{i_m=0}^{t_m} P_{m,i_m}(X_m)\cdot (X_m^q-X_m)^{i_m} \right)\\
&= \mathlarger{\sum}_{\vi = \textbf{0}}^{\vi \leq (t_1,\dots,t_m)} B_{\vi}(X_1,X_2,\dots,X_m)\cdot V_\vi(X_1,X_2,\dots,X_m),
\end{split}
\end{equation*}
where $B_{\vi}(\vX) = P_{1,i_1}(X_1)\cdots P_{m,i_m}(X_m)$ and
$\deg(B_\vi) = \sum_{j=1}^m \deg(P_{j,i_j}) \leq
\min(m(q-1),\sum_{j=1}^m (u_j-qi_j)) = \min(m(q-1), (\sum_{j=1}^m
u_j)-|\vi|q)$. Since a multivariate polynomial is the sum of
multivariate monomials, we obtain the result. We also note that if
there would not exist an $F_{\vj_0}$ such that $|\vj_0| = \lfloor
d/q\rfloor$ and $\deg(F_{\vj_0}) = d-\lfloor d/q\rfloor q$, then the
degree of the RHS of~\refeq{eq:poldecomp} would not be equal to $\deg(F)$.
\end{IEEEproof}
We prove the \emph{uniqueness} of the $F_{\vj}$'s in~\refeq{eq:poldecomp}
in the next Section.
\subsection{Corresponding systematic encoding}

Considering a point $\vP\in V$, we have $V_\vj(\vP+\vZ) =
\sum_\vi \Hasse{V_\vj}{\vi}(\vP) \vZ^\vi$, and,
\begin{equation*}
\begin{split}
  V_\vj(\vP+\vZ) &= \prod_{i = 1}^m ((P_i+Z_i)^q-(P_i+Z_i))^{j_i}\\
  &= \prod_{i = 1}^m \left(Z_i^q-Z_i\right)^{j_i}
  =\vZ^{\vj}\prod_{i=1}^m\left(Z_i^{q-1}-1\right)^{j_i}.
\end{split}
\end{equation*}
So, we have proved the following~\cite{KOP12}:
\begin{equation}\label{Hasse.Deri}
H(V_\vj,\vi)(\vP)=
\begin{cases}
0 & \vi \ngeq \vj\\
(-1)^{|\vi|} & \vi=\vj.\\
\end{cases}
\end{equation}
When we compute the Hasse Derivative of $F$, we find
\begin{align*}
  \Hasse F {\vi}&=\sum_{|\vj| \leq d/q} \Hasse{F_{\vj}V_\vj} \vi\\
  &= \sum_{|\vj| \leq d/q} \sum_{\vu+\vv=\vj}\Hasse{F_\vj}{\vu}\Hasse{V_\vj}{\vv}\\
  \Hasse F {\vi}(\vP) 
&=\sum_{|\vj| \leq d/q}
  \sum_{\vu+\vv=\vj,\vv\leq
    \vj}\Hasse{F_\vj}{\vu}(\vP)\Hasse{V_\vj}{\vv}(\vP).
\end{align*}
Thanks to~\refeq{Hasse.Deri}, the summation  reduces to
\begin{equation}
\begin{split}\label{evalinterset}
  \Hasse F {\vi}(\vP)  &=\sum_{\vj\leq \vi}\;\sum_{\vu+\vv=\vi,\vv\leq \vj}\Hasse{F_\vj}{\vu}(\vP)\Hasse{V_\vj}{\vv}(\vP)\\
  &= (-1)^{|\vi|} F_{\vi}(\vP) +\\
  &\quad \sum_{\vj < \vi}\;\sum_{\vu+\vv=\vi,\vv\leq \vj}\Hasse{F_\vj}{\vu}(\vP)\Hasse{V_\vj}{\vv}(\vP).
\end{split}
\end{equation}
\noindent Thus we can find the evaluation of $F_\vi$ at  $\vP \in \mathbb{F}_q^m$ if we know:
\begin{itemize}
	\item $H(F,\vi)(\vP)$;
	\item the polynomials $F_{\vj}$ for every $\vj < \vi$.
\end{itemize}
Now, using the information set $\Infoset_{{d_\vj}}$ of the Reed-Muller code
$\RM_{d_{\vj}}$ given by Theorem~\ref{theo:infosetRM}, we can
determine $F_\vj$ given the values $F_\vj(\vP)$, $\vP\in
\Infoset_{{d_\vj}}$.  So the set $\Infoset$:
\begin{equation}\label{interset}
\Infoset = (\vj, \Infoset_{d_\vj})_{\vj \in S}
\end{equation}
enables to find $F_\vj$ from its values on $\Infoset_{d_\vj}$. Under
unicity of~\refeq{eq:poldecomp}, we have the following :
\begin{proposition}
An information set of $\MRM_d^s$  is given by~\refeq{interset}.
\end{proposition}

Given a message $M$ of length
$k = \binom{m+d}{m}$ over  $\F_q$, we consider the polynomial $F \in \mathbb{F}_q[\vX]_d$ whose list of coefficients is given by $M$. Then, the classical non-systematic encoding of $M$ is $\ev^s(F) \in \Sigma^n$. 

For the systematic encoding, we write the message as $M =
(M_{\vj,\vP})$, where $\vP \in \Infoset_{d_\vj}$ and $|\vj| \leq d/q$,
and we {\em define} $F$ to be the \emph{unique} polynomial such that
$H(F,\vj)(\vP) = M_{\vj,\vP}$. We then {\em construct} $F$ according
to the above discussion : From the values $H(F,\vj)(\vP)$, we find
$F_\vj$ thanks to~\refeq{evalinterset}. Then we find $F$
using~\refeq{eq:poldecomp} and finally we evaluate $F$ on the remaining
$(\vj,\vP) \notin \mathcal{I}$. The systematic encoding of $M$ over
$V$ is $\ev^s (F)$. We summarize this systematic encoding in
Algorithm~\ref{algo:Sys-MC}.

\begin{algorithm}
\begin{algorithmic}[1]
  \renewcommand{\algorithmicrequire}{\textbf{Input:}} \Require The
  message $M = (M_{\vi,\vP})_{(\vi,\vP)\in \Infoset}$ of dimension
  $k$.  

  \renewcommand{\algorithmicrequire}{\textbf{Output:}} \Require The
  systematic encoding of $M$ over $V$.  \State Determine recursively the polynomials $F_\vj \in \mathbb{F}_q[\vX]$ with
  $|\vj|\leq d/q$, using~\refeq{evalinterset} where $\Hasse{F}{\vi}(\vP)$ is given by
  \[
  \Hasse{F}{\vi}(\vP)=M_{\vi,\vP}, \quad \vi \in S.
  \]
  \State Compute the polynomial $F \in \mathbb{F}_q[\vX]$ as
  $$ F = \sum_{ |\vj| \leq d/q}F_\vj\cdot V_\vj, $$
  where $V_\vj= \prod_{i=1}^m(X_i^q-X_i)^{j_i}$.

  \State\Return $\ev^s(F)$,  the systematic encoding of M over $V$.
\end{algorithmic}
\caption{Systematic encoding algorithm for multiplicity codes}
\label{algo:Sys-MC}
\end{algorithm}
\section{Unicity of the decomposition}
To have unicity of $F$ constructed from the message
$\left(M_{\vj,\vP}\right)_{(\vj,\vP)\in\Infoset}$, and full
correctness of Algorithm~\ref{algo:Sys-MC}, the following statement
suffices.
\begin{lemma}The decomposition~\refeq{eq:poldecomp} in
  Lemma~\ref{lemma:existence} is unique.
\end{lemma}
\begin{IEEEproof}
  \indent We prove this lemma by showing that the size of $\Infoset$
  defined by~(\ref{interset}) is exactly the dimension $k$ of the 
  code. Assume that $d = rq+t$, hence $r \leq s-1$ and $t < q$ (since
  $d < sq$). Recall that the dimension of Reed-Muller codes satisfy $k_{d} =
  |\mathcal{L}_d|=|\Infoset_{d}|$. There are some particular cases:
\begin{itemize}
	\item When $d \geq m(q-1)$, $k_{d} = q^m$
	\item When $0 \leq d \leq q-1$, $k_{d}  = \binom{m+d}{m}$
	\item When $d < 0$, $k_{d} = 0$.
\end{itemize}
Since we do not know any closed formula for $k_{d}$, we use generating
functions (see~\cite{STA11,WIL06}).  First, we give a brief
introduction. If $f(x) = \sum_{n \geq 0} a_nx^n$, then we call $a_n$
the $n$-th coefficient of $x^n$, and denote it by $a_n = [x^n]f(x)$.
Recall that:
\begin{equation}\label{Taylor_k}
\frac{1}{(1-x)^k} = \mathlarger{\mathlarger{\sum}}_{n \geq 0} \binom{n+k-1}{k-1}x^n.
\end{equation}
Using
\begin{equation*}
\begin{split}
\mathcal{K}_{d} =\lbrace ({i_1},\dots,{i_m})\text{ } |\text{ } 0 \leq i_l \leq q-1;\text{ } 1 \leq l \leq m;\text{ }\sum_{l=1}^m i_l = d \rbrace,
\end{split}
\end{equation*}
we have a one-to-one mapping between elements $({i_1},\dots,{i_m})$
$\in \mathcal{K}_{d}$ and monomials $x^{i_1}x^{i_2}\dots x^{i_m}$ of
total degree $d$ and individual degree not greater than $q-1$. Hence,
for a degree $d$, consider the generating function:
\begin{equation*}
\begin{split}
f_m(x) &\triangleq \left( \frac{1-x^q}{1-x} \right)^m\\
&= \underbrace{(1 + x + \cdots + x^{q-1})\cdots (1 + x + \cdots + x^{q-1})}_\text{$m$ times},
\end{split}
\end{equation*}
then the coefficient of $x^d$ of $f_m(x)$ is exactly the cardinality of
$\mathcal{K}_{d}$, with the convention that $\mathcal{K}_{d} =
\emptyset$ when $d > m(q-1)$. From this, we use that $k_d =
|\mathcal{L}_d| = |\mathcal{K}_0| + \cdots + |\mathcal{K}_d|$, with:
\begin{itemize}
	\item[] $|\mathcal{K}_{d}| = [x^d]f_m(x)$,
 $|\mathcal{K}_{d-1}| = [x^{d-1}]f_m(x) = [x^d](xf_m(x))$,
 $\dots,$ $|\mathcal{K}_{0}| = [1]f_m(x) = [x^d](x^d f_m(x))$.
\end{itemize}
Therefore:
\begin{equation*}
\begin{split}
k_{d} & = [x^d](f_m(x) + xf_m(x) + x^2f_m(x) + \cdots + x^df_m(x)) \\
		& = [x^d]\left(\frac{1-x^{D+1}}{1-x}f_m(x)\right)
 		 = [x^d]\left(\frac{f_m(x)}{1-x}\right).
\end{split}
\end{equation*}
\noindent Note that $k_{d} = |\Infoset_d|$. Similarly as above, we have:
\begin{itemize}
	\item[] $k_{d} = [x^d]\frac{f_m(x)}{1-x}$, $k_{d-q} = [x^d]\frac{x^qf_m(x)}{1-x}$, $\dots,$ 
\item[]$k_{d-rq} = [x^d]\frac{x^{rq}f_m(x)}{1-x}$.
\end{itemize}
where $d = rq + t$ and $t <q$. For every $\vj$ we have $\binom{m-1+|\vj|}{m-1}$ such sets $(\vj,\Infoset_{d_\vj})$. By~\refeq{interset}, it follows that the size of $\Infoset$ is thus
\begin{equation}\label{eq:recurrence}
|\Infoset| = \sum_{u = 0}^{s-1} \sum_{|\vj| = u} |\Infoset_{d_\vj}| = \mathlarger{\mathlarger{\sum}}_{j=0}^r \binom{m-1+j}{m-1} k_{d-jq},
\end{equation}
which implies
\begin{equation*}
\begin{split}
|\Infoset| 	& = [x^d]\frac{f_m(x)}{1-x} + \binom{m-1+1}{m-1}\cdot [x^d]\left( \frac{x^q}{1-x}f_m(x) \right) \\
		& \text{    } + \cdots + \binom{m-1+r}{m-1}\cdot [x^d]\left( \frac{x^{rq}}{1-x}f_m(x) \right)\\
 		& = [x^d]\left(\frac{ \sum_{i=0}^r \binom{m-1+i}{m-1}x^{iq} }{1-x}f_m(x)\right).
\end{split}
\end{equation*}
Using~\refeq{Taylor_k}, we have:
$$
\mathlarger{\mathlarger{\sum}}_{i\geq 0} \binom{m-1+i}{m-1}x^{iq} = \frac{1}{(1-x^q)^m},\:\mbox{ so}
$$
\begin{equation*}
\begin{split}
|\Infoset| 	& = [x^d]\left(\frac{ \sum_{i=0}^r \binom{m-1+i}{m-1}x^{iq} }{1-x}f_m(x)\right) \\
		& = [x^d]\left(\frac{ \sum_{i\geq 0} \binom{m-1+i}{m-1}x^{iq} }{1-x}f_m(x)\right)\\
		& = [x^d]\left(\frac{1}{(1-x^q)^m(1-x)}f_m(x)\right)\\
		& = [x^d]\left(\frac{1}{(1-x^q)^m(1-x)} \left( \frac{1-x^q}{1-x} \right)^m \right)\\
		& = [x^d]\left(\frac{1}{(1-x)^{m+1}} \right)\\
		& = \binom{m+d}{m} = k,
\end{split}
\end{equation*}
as we wanted to prove.
To conclude the proof, we consider
$$
\psi:\begin{array}[t]{rcl}
\prod_{|\vj| \leq d/q}\mathbb{F}_q[\vX]_{d_{\vj}} &\rightarrow &\quad \mathbb{F}_q[\vX]_d\\
(F_\vj)_{|\vj|\leq d/q} & \mapsto & \sum_{|\vj|\leq d/q} F_\vj\cdot V_\vj = F
\end{array}
$$
Lemma 1 shows that $\phi$ is surjective. Since we have just proved
$$
\dim(\mathbb{F}_q[\vX]_d) = \binom{m+d}{m} = \sum_{|\vj| \leq d/q} \dim(\mathbb{F}_q[\vX]_{d_\vj}),
$$
the equality of dimensions of the range and of the domain implies that
$\psi$ is bijective, in particular one-to-one.
\end{IEEEproof}

Note that from Equation~\refeq{eq:recurrence}, we can compute easily
the value of $k_{d}$ recursively from $k_{d-iq}$'s where $0\leq i\leq
d/q$.
\section{Systematic encoding for Derivative Codes}
In this Section, we apply the previous results to the particular
 case of $m = 1$. This boils down to codes generalizing
Reed-Solomon codes, using derivatives. These codes have been used
in~\cite{GW13}, where they were given the name of \emph{Derivative
  Codes}.  Let be given $s$ and $d$ as in
Definition~\ref{def:mult}. In this case, the information sets
$\Infoset_{d_j}$ are
$$
\Infoset_{d_j} = \left\{ i\text{ } |\text{ } 0 \leq i \leq d_j \right\},\quad j =0,\dots, s-1.
$$
The systematic encoding is described in Algorithm~\ref{algo:Sys-MRS}.
\begin{algorithm}
\begin{algorithmic}[1]
\renewcommand{\algorithmicrequire}{\textbf{Input:}}
\Require The message $M = (M_{i,P})$ of dimension $k$, where $P \in \mathcal I_{d_i}$ and $i < s$.
\renewcommand{\algorithmicrequire}{\textbf{Output:}}
\Require The systematic encoding of $M$ over $\mathbb{F}_q$.
  \State Find the polynomials $F_i \in \mathbb{F}_q[X]$ where $i < s$, such that:
  \begin{equation*}
  \begin{split}
  &M_{i,P} = (-1)^{i} F_{i}(P)+\\
  &+\sum_{j = 0}^{i-1}\;\sum_{v = j}^i H(F_{j},i-v)(P)H(V_{j},v)(P)
  \end{split}
  \end{equation*}  
  \State Define the polynomial $F \in \mathbb{F}_q[X]$ as
  $$ F = \sum_{j<s}F_{j}\cdot V_{j}, $$
  where $V_{j} = (X^q-X)^{j}$.
  \State\Return $\ev(F)$,  the systematic encoding of $M$ over $\mathbb{F}_q$.
\end{algorithmic}
\caption{Systematic encoding algorithm for Derivative codes}
\label{algo:Sys-MRS}
\end{algorithm}



\section{Conclusion}
We have defined the notion of information set for Multiplicity codes
as $\mathbb{F}_q$-linear codes. We filled in details of the work of
Kopparty~\cite{KOP12}, who introduced a systematic encoding for such
codes. Our work also allowed us to propose a new recursive formula for
the size of Reed-Muller codes over $\mathbb{F}_q$, that makes use of a
combinatorial proof of generating functions. Designing efficient
algorithms for fast systematic encoding will be the topic of future
work.

\section{Acknowledgment}
The third author would like to thank Doron Zeilberger and Louis Joseph Billera for the suggestion of using generating functions in Section IV.
\bibliographystyle{IEEEtran}
\bibliography{ieee}

\begin{thebibliography}{10}
\providecommand{\url}[1]{#1}
\csname url@samestyle\endcsname
\providecommand{\newblock}{\relax}
\providecommand{\bibinfo}[2]{#2}
\providecommand{\BIBentrySTDinterwordspacing}{\spaceskip=0pt\relax}
\providecommand{\BIBentryALTinterwordstretchfactor}{4}
\providecommand{\BIBentryALTinterwordspacing}{\spaceskip=\fontdimen2\font plus
\BIBentryALTinterwordstretchfactor\fontdimen3\font minus
  \fontdimen4\font\relax}
\providecommand{\BIBforeignlanguage}[2]{{%
\expandafter\ifx\csname l@#1\endcsname\relax
\typeout{** WARNING: IEEEtran.bst: No hyphenation pattern has been}%
\typeout{** loaded for the language `#1'. Using the pattern for}%
\typeout{** the default language instead.}%
\else
\language=\csname l@#1\endcsname
\fi
#2}}
\providecommand{\BIBdecl}{\relax}
\BIBdecl

\bibitem{KT00}
J.~Katz and L.~Trevisan, ``On the {E}fficiency of {L}ocal {D}ecoding
  {P}rocedures for {E}rror-correcting {C}odes,'' in \emph{Proceedings of the
  Thirty-second Annual ACM Symposium on Theory of Computing, STOC '00}, F.~Yao
  and E.~Luks, Eds.\hskip 1em plus 0.5em minus 0.4em\relax ACM, 2000, pp.
  80--86.

\bibitem{YEK10}
S.~Yekhanin, \emph{Locally Decodable Codes}, ser. Foundations and Trends in
  Theoretical Computer Science.\hskip 1em plus 0.5em minus 0.4em\relax NOW
  publisher, 2012, vol.~6.

\bibitem{KSY11}
S.~Kopparty, S.~Saraf, and S.~Yekhanin, ``High-rate {C}odes with
  {S}ublinear-time decoding,'' in \emph{Proceedings of the Forty-third Annual
  ACM Symposium on Theory of Computing, STOC'11}, S.~Vadhan, Ed.\hskip 1em plus
  0.5em minus 0.4em\relax New York, USA: ACM, 2011, pp. 167--176.

\bibitem{AFS2014}
D.~Augot, F.~Levy{-}dit{-}Vehel, and A.~Shikfa, ``A storage-efficient and
  robust private information retrieval scheme allowing few servers,'' in
  \emph{Cryptology and Network Security - 13th International Conference, {CANS}
  2014, Heraklion, Crete}, ser. Lecture Notes in Computer Science.\hskip 1em
  plus 0.5em minus 0.4em\relax Springer, 2014, pp. 222--239.

\bibitem{KOP12}
S.~Kopparty, ``List-decoding multiplicity codes,'' \emph{Electronic Colloquium
  on Computational Complexity (ECCC)}, vol. TR12-044, 2012.

\bibitem{KLP68}
T.~Kasami, S.~Lin, and W.~Peterson, ``New generalizations of the
  {R}eed-{M}uller codes. {I}. {P}rimitive codes,'' \emph{IEEE Trans.
  Information Theory}, vol.~14, no.~2, pp. 189--199, 1968.

\bibitem{KMM06}
J.~Key, T.~McDonough, and V.~Mavron, ``Information sets and partial permutation
  decoding for codes from finite geometries,'' \emph{Finite Fields and Their
  Applications}, vol.~12, no.~2, pp. 232--247, Apr. 2006.

\bibitem{STA11}
R.~P. Stanley, \emph{Enumerative Combinatorics}, ser. Cambridge Studies in
  Advanced Mathematics.\hskip 1em plus 0.5em minus 0.4em\relax Cambridge
  University Press, 2011, vol.~1.

\bibitem{WIL06}
H.~S. Wilf, \emph{Generatingfunctionology}.\hskip 1em plus 0.5em minus
  0.4em\relax A. K. Peters, Ltd. Natick, 2006.

\bibitem{GW13}
V.~Guruswami and C.~Wang, ``Linear-algebraic list decoding for variants of
  {R}eed--{S}olomon codes,'' \emph{Information Theory, IEEE Transactions on},
  vol.~59, no.~6, pp. 3257--3268, Jun. 2013.

\end{thebibliography}

\appendix[Complexity estimates]
We give a rough and conservative estimate on the number of arithmetic
operations in $\F_q$ needed for systematic encoding. Algorithm~\ref{algo:Sys-MC} finds a unique
polynomial $F\in\F_q[\vX]$ from the
$F_{\vi}$'s, those $F_{\vi}$'s being found from the $\F_q$-symbols $M_{\vj,\vP}$ at the $(\vj,
\vP)\in\Infoset=\left(\vj,\Infoset_{d_\vj}\right)_{\vj\in S}$; then it 
evaluates back this polynomial $F$ for $(\vj,\vP)\not\in \Infoset$.
But~\refeq{eq:poldecomp} requires expensive multiplications of
multivariate polynomials.  Yet~\refeq{evalinterset} also enables to
bypass the computation of $F$, working only with $F_{\vj}$'s, as follows.
At step $\vi$, a first pass consists in going through the points 
$\vP\in\Infoset_{d_\vi}$ to compute $F_\vi(\vP)$. Then
$F_\vi\in\F_q[\vX]_{d_{\vi}}$ is uniquely determined by its values on the information set
$\Infoset_{d_\vi}$. Note that $F_\vi$ can be computed by applying the
(precomputed) inverse of $\varphi$ defined in~\refeq{eq:mapinfoset},
i.e.\ a matrix-vector product of cost $O(k_{d_\vi}^2)$. Once $F_\vi$
is computed, using~\refeq{evalinterset} again, the values
$\Hasse{F}{\vi}(\vP)$, for $\vP\not\in\Infoset_{d_{\vi}}$ are computed.
With $\sigma=\size{S}$, we have, for each $\vi\in S$:
\begin{enumerate}
\item for each $\vP\in\Infoset_{d_\vi}$, $O(\sigma^2)$ for computing
 $F_{\vi}(\vP)$ using~\refeq{evalinterset}; 
  thus a total of $k_{d_{\vi}}\sigma^2$ for all  $\vP\in\Infoset_{d_\vi}$;
\item $O(k_{d_\vi}^2)$ for recovering $F_\vi$, using a
  matrix-vector product;
\item $O(\sigma k_{d_\vi})$ for computing the $\sigma $ Hasse
  derivatives of $F_\vi$, (termwise on $F_\vi$, step-by-step through
  $S$);
\item  $\tilde O(n)$ for at once evaluating $F_\vi$ on all $P\not
  \in \Infoset_{d_\vi}$, neglecting logarithmic factors (multidimensionnal FFT)
\item for each $\vP\not\in\Infoset_{d_\vi}$, $O(\sigma^2)$ for
  computing each $\Hasse{F}{\vi}(\vP)$ using~\refeq{evalinterset}
  again, for a total of $(n-k_{d_{\vi}})\sigma^2$.
\end{enumerate}
Summing over the $\vi\in S$, we get a ``soft-$O$'' estimate of 
$ \tilde
O\left(\sum_{\vi\in S}n\sigma^2+k_{d_\vi}^2\right)=\tilde
O\left(n\sigma^3+k^2\right) $, with a memory footprint of $O(\sigma n) $ for storing all
the $F_\vi$'s and their Hasse derivatives. Note that $\sigma n$ is the size
of the output codeword.
\end{document}